\documentclass[5p,twocolumn]{elsarticle}
\journal{Physics Letters B}

\usepackage{amsmath,amssymb,mathrsfs,bm}
\usepackage{url}
\usepackage{hyperref}
\usepackage{textcomp}
\hypersetup{
colorlinks=true,
linkcolor=blue,
anchorcolor =red,
citecolor=blue,
filecolor = red,
urlcolor=blue,
pdfauthor=author}
\usepackage{txfonts}
\usepackage[utf8]{inputenc}
\usepackage{comment}
\def\av<#1>{\left\langle\,#1\,\right\rangle}
\def\ev<#1>{\left\langle\,#1\,\right\rangle_{\rm{ev}}}

\begin{document}

\begin{frontmatter}

\title{String breaking and running coupling of $\rm Q\bar{Q}$ in a rotating media from holography}

\author[1,2,3]{Jing Zhou}

\author[2,3]{Saiwen Zhang}

\author[4]{Jun Chen}

\author[5]{Le Zhang}

\author[1]{Xun Chen\corref{cor}}
\ead{chenxunhep@qq.com}
\cortext[cor]{Corresponding author}

\affiliation[1]{organization={School of Nuclear Science and Technology, University of South China},
	city={Hengyang},
	postcode={421001},
	country={China}}
\affiliation[2]{organization={Department of Physics, Hunan City University},
	city={Yiyang},
	postcode={413000},
	country={China}}
\affiliation[3]{organization={All-solid-state Energy Storage Materials and Devices Key Laboratory of Hunan Province, Hunan City University},
	city={Yiyang},
	postcode={413000},
	country={China}}
\affiliation[4]{organization={Department of Physics, Hubei Minzu University},
	city={Enshi},
	postcode={445000},
	country={China}}
\affiliation[5]{organization={College of Physics and Electronic Science, Hubei Normal University},
	city={Huangshi},
	postcode={435002},
	country={China}}

\begin{abstract}
Through gravity/gauge duality, the string breaking and running coupling constant of heavy quark-antiquark pair are investigated in the rotating background. For the meson decay mode $\rm Q\bar{Q} \rightarrow Q \bar{q}+\bar{Q} q$, we discuss the string breaking and running coupling in the parallel and transverse case. It is found that the parallel case has a more significant on string breaking and running coupling constant than the transverse case in the confined phase. The string-breaking distance and the maximum value of running coupling will decrease with the increase of angular velocity in the parallel case. Besides, we also investigate the running coupling and screening distance at finite angular velocity in the deconfined phase. It is found that the maximum values of the running coupling and screening distance are decreasing functions of angular velocity. The parallel case has a more significant influence on the running coupling and screening distance than the transverse case in the deconfined phase.
\end{abstract}


\end{frontmatter}

\section{Introduction}\label{sec:intro}
Quantum chromodynamics(QCD) has two specific features, namely, asymptotic freedom \cite{Gross:1973id,Gross:1973ju,Politzer:1974fr} and quark confinement \cite{Wilson:1974sk,Weinberg:1973un}. The former means that the coupling constant decreases with the increase of the transferred momentum. The latter means that there is no single free quark, namely, quarks are bound inside hadrons. The researches on these two features can deepen our understanding of QCD properties. In QCD, the running coupling constant plays key role in the physics many important processes, such as jet quenching, heavy quarkonium production and Higgs production \cite{CarcamoHernandez:2013ydh}. Thus, it is necessary to know how the running coupling constant changes from short distance to long distance. The running coupling constant has been widely investigated in Refs.\cite{Javidan:2020lup,Giunti:1991ta,Baikov:2016tgj,Aguilar:2001zy,Lombardo:1996gp,Lombardo:1996gp,Bloch:2003sk,Deur:2016tte,Yu:2021yvw,Takaura:2018vcy}. As we know, extreme conditions(high temperature, high density, strong magnetic field, large angular momentum) will be formed in heavy-ion collision. Refs.\cite{Kaczmarek:2004gv,Kaczmarek:2005ui} discussed the running coupling constant at finite temperature through lattice QCD. The effect of chemical potential and magnetic field on running coupling constant was studied in Ref.\cite{Chen:2021gop} and we found that the coupling constant is suppressed by chemical potential and magnetic field.

Heavy quarkonium is an important probe which can help us to understand the property of quark-gluon plasma(QGP). In Ref.\cite{Maldacena:1998im}, Maldacena first proposed the calculation of heavy quark-anqiquark potential in holographic way. Then, the proposal of Maldacena was clearly explained and generalized in \cite{Kinar:1998vq}. In the holographic model, heavy quarkonium is connected by a string which tends to infinity in confined phase. However, due to the existence of dynamic quarks or sea quarks, the string will break in a limited distance \cite{Philipsen:1998de}. In addition, the decay mode $\rm Q \bar{Q} \rightarrow Q \bar{q}+\bar{Q} q$ can be regarded as an important mechanism for $\rm Q \bar{q}$ or $\rm \bar{Q} q$ generation.  Although lattice QCD has achieved some important results in the study of string breaking\cite{Yamamoto:2008jz}, this method has its limitations. Besides, A lot of works related to string breaking have been carried out in recent years\cite{Chen:2021bkc,Bulava:2019iut,Andreev:2021eyj,Bruschini:2021fuw,Andreev:2021vjr,Andreev:2021bfg,Astrakhantsev:2018uzd,Simonov:2000hd,Knechtli:1998gf,Drummond:1998qb,Okiharu:2004ve,Cardoso:2011fq}.

In non-central relativistic heavy-ion collisions, the system will produce non-zero angular momentum and some angular momentum will remain in QGP. Recently, many researches have been done to study the rotational effect of QGP \cite{Jiang:2016woz,Jiang:2015cva,McInnes:2014haa,Liang:2019clf,Zhou:2021sdy,Tuchin:2021lxl,Huang:2020xyr,Wang:2018sur,Liu:2017spl,Huang:2017pqe}. The dynamics of a heavy quark in a strongly coupled rotating quark-gluon plasma within the holographic framework is explored in Ref.\cite{Golubtsova:2021agl}. Confinement-deconfinement phase transition under the rotating quark-gluon plasma was discussed in Refs. \cite{Chen:2020ath,Braga:2022yfe}. The drag force in rotating quark-gluon plasma was studied in Refs. \cite{Arefeva:2020knc,Arefeva:2020jvo}. The unique advantage of gauge/gravity duality \cite{Maldacena:1997re,Witten:1998qj,Gubser:1998bc} is that it can be used to study the strongly coupled QGP from the perspective of gravity\cite{Chen:2017lsf,Chen:2019rez,Jiang:2022zbt,Cao:2022csq,Cao:2022mep,Zhang:2023kzf,Zhang:2023psy,Tahery:2022pzn,Wu:2022ufk,Cai:2022nwq,Contreras:2021epz,Zhao:2022uxc}.

In this paper, we mainly study string breaking and running coupling in the 5-dimensional rotating deformed $\rm AdS_5$ black hole\cite{Zhou:2021sdy}. The rest parts of the paper are organized as follows. In Sec.~\ref{sec:02}, we briefly review the introduction of the deformed
rotating $\rm AdS_5$ black hole. In Sec.~\ref{sec:03}, we mainly focus on the string breaking and running coupling constant at finite angular velocity in confined phase. In Sec.~\ref{sec:04}, the string breaking and running coupling constant are discussed at finite angular velocity in the deconfined phase. Further discussion and conclusion can be found in Sec.~\ref{sec:05}.
\vspace{10pt}
\section{The Setup}\label{sec:02}
Following Ref.\cite{Zhou:2021sdy}, we start with the deformed $\rm AdS_5$ black hole and then extend to the rotating case. Despite the deformed $\rm AdS_5$ metric is not a self-consistent solution of Einstein equation, we can still gain some insights into problems for which there are no predictions from phenomenology and the lattice. The static metric is\cite{Andreev:2006eh}
\begin{eqnarray}
\mathrm{d} s^{2} & = & \frac{L_{\text {AdS}}^{2} h(z)}{z^{2}}\left(-f(z) \mathrm{d} t^{2}+\sum\limits_{i=1}^3dx_i^2 + \frac{\mathrm{d} z^{2}}{f(z)}\right), \\
f(z) & = & 1-\frac{z^{4}}{z_{h}^{4}} .
\end{eqnarray}
Here $h(z) = \exp(cz^2/2)$ is the warp factor, where $c$ determines the deviation from conformality. Here we take $c = 0.9 \mathrm{GeV}^{2}$ which is determined by $\rho$ meson trajector\cite{Andreev:2006vy} and $L_{AdS}=1$. $z_h$ is position of black hole horizon, $x_i$ is the three-dimensional coordinate and $z$ is the fifth coordinate.
Following \cite{Zhou:2020ssi,BravoGaete:2017dso,Nadi:2019bqu,Erices:2017izj,Chen:2020ath,Zhao:2022uxc}, we can extend the deformed $\rm AdS_5$ black hole to the rotating case. To introduce the rotation effect, the deformed $\rm AdS_5$ metric can be rewritten as
\begin{eqnarray}
\mathrm{d} s^{2} & = & \frac{ h(z)}{z^{2}}\left(-f(z) \mathrm{d} t^{2}+\mathrm{d} x_1^2+ \mathrm{d} x_2^2+l^2\mathrm{d} \phi^2+\frac{\mathrm{d} z^{2}}{f(z)}\right),
\end{eqnarray}
where $\phi$ denotes the angle in cylindrical coordinates. The angular velocity is introduced through the standard Lorentz transformation
\begin{equation}
\begin{aligned}
t&\rightarrow \frac{1}{\sqrt{1-\omega^2 l^2}}(t+\omega l^2 \phi), \\
\phi&\rightarrow  \frac{1}{\sqrt{1-\omega^2 l^2}}(\phi+\omega t).
\end{aligned}
\end{equation}
Here $l$ is the rotating radius and $\omega$ is the angular velocity. The strong interacting
matter produced by ultra relativistic heavy ion collisions is finite in volume, and its size depends on the nature
of the colliding nuclei, the center of mass energy and the centrality of collision\cite{Zhao:2018nqa}. In Ref.\cite{Zhang:2001vk}, the author estimates the size of QGP at RHIC and LHC will be around 4-8 fm and 6-11 fm respectively. Thus, it is reasonable for us to fix the rotating radius as $l=10\rm{GeV^{-1}} \sim 2 \rm{fm}$. Next we extend the deformed $\rm AdS_5$ black hole to the rotating black hole with planar horizon. The general metric is
\begin{equation}
ds^2 = g_{tt}dt^2+g_{t\phi}dtd\phi+g_{\phi t}d\phi dt+g_{\phi\phi}l^2d\phi^2+g_{xx}\sum\limits_{i=1}^2dx_i^2,
\end{equation}
and in our case
\begin{equation}
\begin{aligned}
g_{tt}& = H(z)\gamma^2(\omega^2 l^2-f(z)), \\
g_{\phi\phi}&=H(z)\gamma^2(1-\omega^2f(z)l^2), \\
g_{t\phi}&=g_{\phi t}=H(z)\gamma^2\omega(1-f(z))l^2,\\
g_{zz}&=\frac{H(z)}{f(z)},\\
g_{xx}&=H(z),\\
\gamma& = \frac{1}{\sqrt{1-\omega^2 l^2}}.
\end{aligned}
\end{equation}
Next, the metric in rotating case can be rewritten as
\begin{equation}
ds^2 = -N(z)dt^2 + \frac{H(z)dz^2}{F(z)} + R(z) (d\phi + P(z)dt)^2 + H(z)\sum\limits_{i=1}^2dx_i^2,
\end{equation}
where
\begin{equation}
\begin{aligned}
N(z)&= \frac{H(z)f(z)(1-\omega^2 l^2)}{1-f(z)\omega^2 l^2}, \\
R(z)&= H(z)\gamma^2 l^2 - H(z)f(z)\gamma^2\omega^2 l^4,  \\
P(z)&= \frac{\omega - f(z)\omega}{1 - f(z)\omega^2 l^2}.\\
H(z)&= \frac{h(z)}{z^2}. \\
\end{aligned}
\end{equation}
Then, the Hawking temperature in our case can be calculated as\cite{Awad:2002cz,Zhou:2021sdy}
\begin{equation}
T = \frac{1}{ \pi z_h} \sqrt{1-\omega^{2}}.
\end{equation}
The Nambu-Goto action of the string is defined as
\begin{equation}
S_{NG} = - \frac{1}{2\pi\alpha'}\int d^2{\xi} \sqrt{- \det g_{ab}},
\end{equation}
where $g_{ab}$ is the induced metric
\begin{equation}
g_{ab} = g_{MN} \partial_a X^M \partial_b X^N, \quad a,\,b=0,\,1.
\end{equation}
Here $X^M$ are the coordinates and $g_{MN}$ is the metric of the AdS space. We define $\mathbf{g}=\frac{1}{2 \pi \alpha^{\prime}}$ and consider the the static gauge $\xi^0 = t,\,\xi^1 = x_2$ or $x_3(\phi)$.
Thus, the Nambu-Goto action can be rewritten as
\begin{equation}
S_{N G}  = -\frac{\mathbf{g}}{T} \int_{-L / 2}^{L / 2} \mathrm{d} \xi_{1} \sqrt{k_{1}(z) \frac{\mathrm{d} z^{2}}{\mathrm{d} \xi^{2}}+k_{2}(z)}.\label{Snng}
\end{equation}
\begin{figure}
	\resizebox{0.45\textwidth}{!}{
		\includegraphics{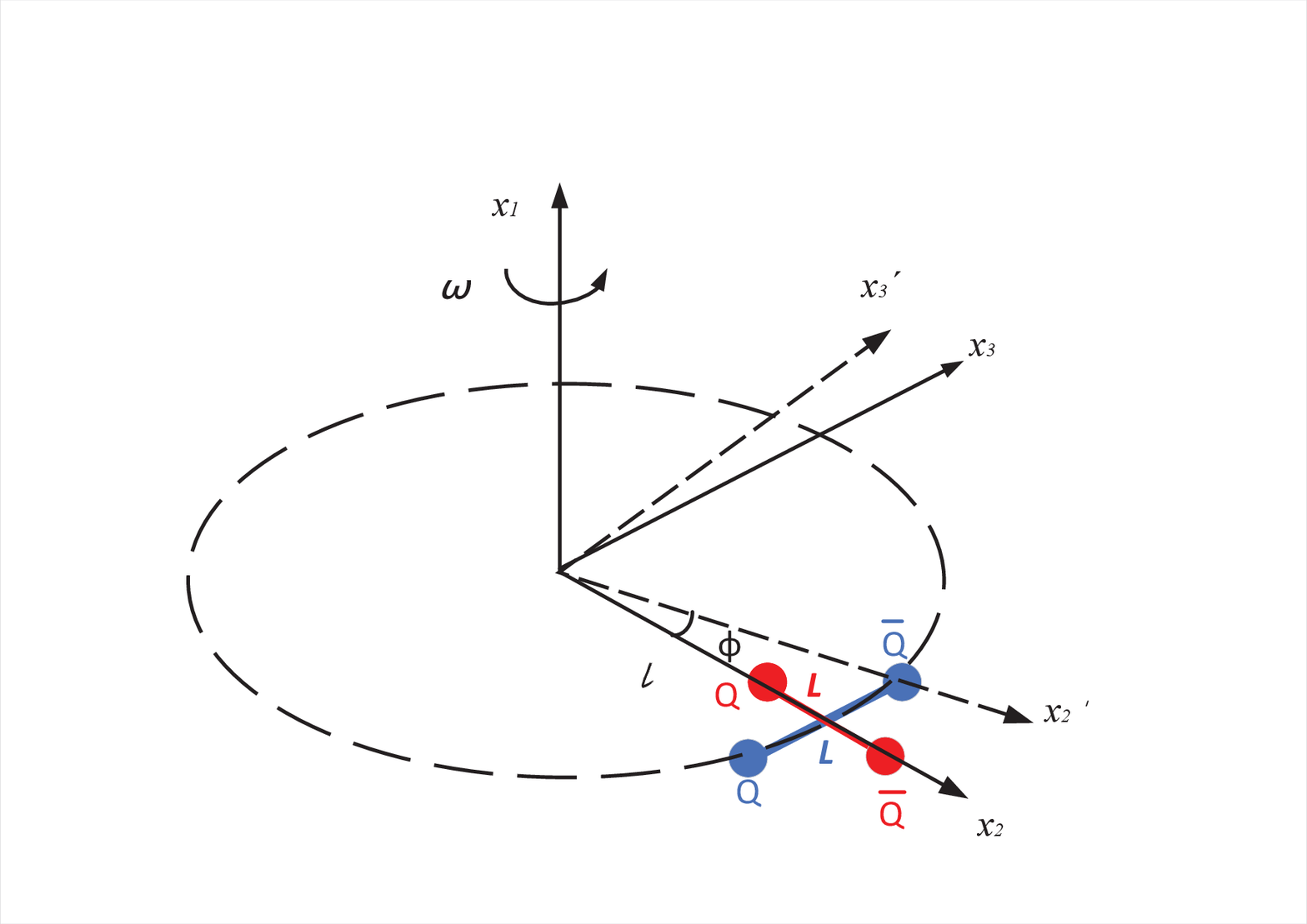}}
	\caption{\label{1} A schematic picture of $\rm Q\bar{Q}$ in the rotating background. The blue quark-antiquark pair is defined as parallel case. The red quark-antiquark pair is defined as transverse case.}
\end{figure}
First, the quark-antiquark pair is put on the position ($x_1= 0 , x_2 = l,  x_3 = 0$) in the Cartesian coordinate. When the quark-antiquark pair is put along the $x_2$ direction as shown in Fig.~\ref{1} with red color, we have
\begin{gather}
k_1(z)=k^{tra}_1(z) =\frac{(N(z)-R(z)P(z)^2)H(z)}{f(z)},\\
k_2(z)=k^{tra}_2(z) =-(R(z)P(z)^2- N(z))H(z).
\end{gather}
We define this as the transverse case with the following boundary conditions:
\begin{gather}
z(x_2=l \pm \frac{L}{2})=0,z(x_2=l)=z_0,z'(x_2=l)=0.
\end{gather}
For the quark-antiquark pair put along the $x_3$ direction with blue color, we have $k_1^{\mathrm{par}}(z)=k^{tra}_1(z)$ and $k_2(z)=k_2^{\mathrm{par}}(z)=N(z) R(z)$ with boundary conditions
\begin{gather}
z(x_3=\pm \frac{L}{2})=0,z(x_3=0)=z_0,z'(x_3=0)=0,
\end{gather}
which is the parallel case. The separate distance of quark-antiquark pair can be calculated as
\begin{equation}
L = 2 \int_0^{z_0} [\frac{k_2(z)}{k_1(z)} (\frac{k_2(z)}{k_2(z_0)}-1)]^{-1/2} dz.\label{distance}
\end{equation}
The renormalized potential energy(free energy) of $\rm Q\bar{Q}$ pair can be calculated as
\begin{equation}
\begin{split}
E_{Q\bar{Q}} = 2 \mathbf{g} &\int_0^{z_0} d{z}(\sqrt{\frac{k_2(z)k_1(z)}{k_2(z)-k_2(z_0)}} - \sqrt{k_2(z\rightarrow 0)})\\ &- 2 \mathbf{g} \int_{z_0}^\infty \sqrt{k_2(z\rightarrow 0)} d{z}. \label{freeenergy}
\end{split}
\end{equation}
Note that we subtract the divergent part an appropriate (infinite) quantity to get the finite potential energy. The detailed detailed analysis can be found in Ref.\cite{Ewerz:2016zsx}. Then, the energy of the single heavy quark can be considered as large-distance limit of
the free energy of heavy $\rm Q\bar{Q}$
\begin{equation}
E_{Q} = \mathbf{g}(\int_0^{z_h} dz (\sqrt{k_1(z)}-\frac{1}{z^2})-\frac{1}{z_h}).
\end{equation}

At the end of this section, we will give a discussion of the physical image. Since gluon dynamic can be well described by the deformed $\rm AdS_5$ metric, we extend the gluon background to the rotating case and put a static heavy quark-antiquark pair into this background. Besides, someone may argue $\omega*l$ goes obviously beyond the speed of light when $\omega$ is increased from zero to pretty large value. Actually, we only consider a finite-size system with a fixed $l$. Moreover, the whole system is obviously a fluid instead of strictly rigid body in the real QGP. Certainly, we can consider $\omega$ is $l$ dependent as done in Ref.~\cite{Chen:2022mhf}. In our work, we assume the influence of rotation on the $\rm Q\bar{Q}$ with the same angular velocity and ignore the size of $\rm Q\bar{Q}$. In addition, the influence of rotation depends on the spatial location. Note that we only focus on the fixed radius in this paper.

\vspace{10pt}
\section{String breaking and running coupling constant in confined phase}\label{sec:03}
As a extension of our previous works\cite{Zhou:2021sdy,Chen:2021gop,Chen:2021bkc}, we continue to calculate the the string breaking and running coupling of $\rm Q\bar{Q}$ in the rotating background. In this paper, we only focus on the meson decay mode
\begin{eqnarray}
\rm Q \bar{Q} \rightarrow Q \bar{q}+\bar{Q} q.
\end{eqnarray}
The Nambu-Goto action of $\rm Q \bar{Q}$ has been given by Eq.(\ref{Snng}). Now the main problem is to calculate the action of $\rm Q \bar{q}$. In the background of open-string tachyon, the constant tachyon field $\mathrm{T}(\mathrm{x}, \mathrm{r})$ is assumed to be point-like particle with mass in five-dimensional AdS space~\cite{Andreev:2015riv}. Assuming $\mathrm{T}(\mathrm{x}, \mathrm{r})=\mathrm{T}_{0}$, the action of the light quark can be computed through tachyon field
\begin{eqnarray}
S_{\mathrm{q}} & = & \mathrm{m} \int d t  \sqrt{N(z)-R(z)P(z)^{2}},
\end{eqnarray}
where $\mathrm{m}= \mathrm{~T}_{0}$ and $\mathrm{~T}_{0}$ is the current quark mass of light flavor. The mass of the light quark is roughly estimated as 46.6MeV through string breaking in Ref. \cite{Andreev:2019cbc}. We redefine $\mathbf{g}=\frac{R^{2}}{2 \pi \alpha^{\prime}}$ and $\mathrm{n}=\frac{\mathrm{m}}{\mathbf{g}}$. Therefore, the total action of $\rm Q \bar{q}$ is the Nambu-Goto action plus the light quark action
\begin{eqnarray}
S_{Q \bar{q}}= S_{\mathrm{NG}}+S_{\mathrm{\bar{q}}}.
\end{eqnarray}
Here $z_{q}$ is the position of light quark which can be obtained by solving $\delta S_{Q \bar{q}}=0$, namely,
\begin{equation}
\begin{aligned}
\sqrt{k_1(z)}+n \left(\sqrt{N(z) -R(z_{q}) \mathrm{P}(z_{q})^{2}}\right)^{\prime}=0.
\end{aligned}
\end{equation}
The prime denotes the derivative with respect to $z_q$. The model parameters are taken as follows: we obtain $\mathbf{g}= 0.176$ by fitting the string tension and $n= 3.057$\cite{Bulava:2019iut,Andreev:2019cbc}. With the position of light quark $z_{q}$ in hand, the energy of $Q \bar{q}$ can be calculated as
\begin{equation}
\begin{aligned}
E_{Q \bar{q}} = & \mathbf{g}\left(\int_{0}^{z_{q}} \left(\sqrt{\frac{\left(N(z)-R(z) P(z)^{2}\right) H(z)}{f(z)}} -\frac{1}{z^{2}}\right) dz \right)-\frac{\mathbf{g}}{z_{q}}\\
&+\mathbf{g} n \sqrt{\frac{\mathrm{H}(z_{q}) \mathrm{f}(z_{q}) \left(1-\omega^{2}\right)}{1-f(z_{q})  \omega^{2}}-R(z_{q}) (\mathrm{P}(z_{q}))^{2}}.
\end{aligned}
\end{equation}
Next, we need to calculate the energy of $\rm Q\bar{Q}$, namely,
\begin{eqnarray}
E_{Q \bar{Q}} & = & 2 \mathbf{g}\left(\int_{0}^{z_{0}} \sqrt{\frac{k_{2}(z) k_{1}(z)}{k_{2}(z)-k_{2}\left(z_{0}\right)}}-\frac{1}{z^{2}}\right)-\frac{2\mathbf{g} }{z_{0}}.
\end{eqnarray}
With above two formulas in hand, we can calculate the string-breaking distance which happens at $2E_{Q \bar{q}} = E_{Q \bar{Q}}(L_c)$.  After deriving the formulas of string breaking, we turn to study the running coupling constant at finite angular velocity. Motivated by perturbation theory for free energy, one can define the running coupling constant for the singlet free energy from lattice QCD as \cite{Kaczmarek:2005ui}
\begin{eqnarray}
\alpha_{\mathrm{Q} \bar{Q}} & = & \frac{3 L^{2}}{4} \frac{\mathrm{d} E_{Q \bar{Q}}}{\mathrm{~d} L}.
\end{eqnarray}
Here $L$ is the separate distance of heavy quark-antiquark, which is given by Eq.(\ref{distance}). The undetermined constant contribution to the heavy quark potential cancels out in this scheme. Moreover, the large distance, non-perturbative confinement contribution to $\alpha_{q q}(r)$ is positive and allows for a smooth matching of the perturbative short distance coupling to the non-perturbative large distance confinement signal. In any case, however, in the non-perturbative regime the value of the coupling will depend on the observable used for its definition\cite{Kaczmarek:2005ui}.
\begin{figure*}
	\centering
	\includegraphics[width=17cm]{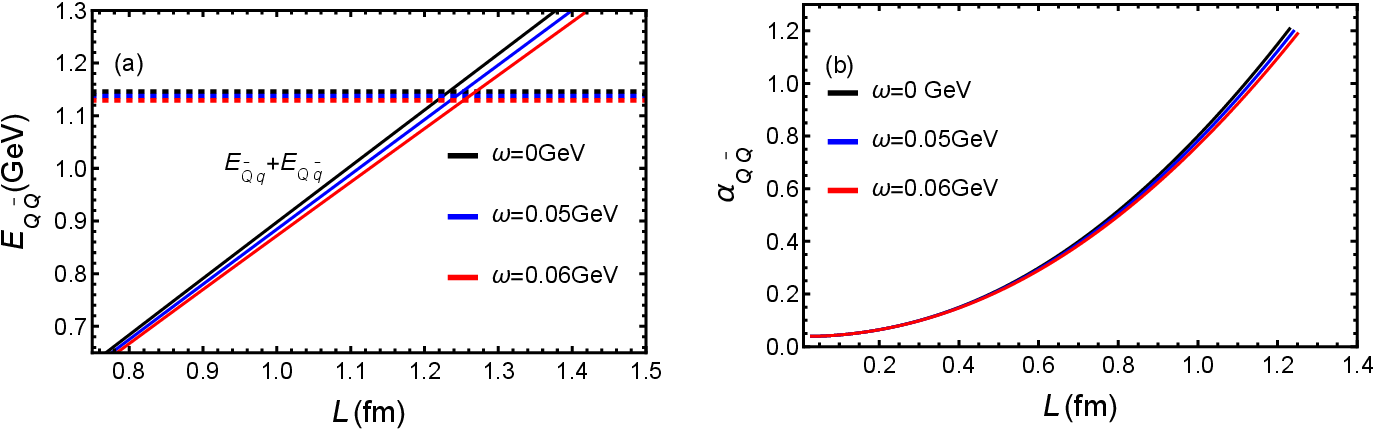}
	\caption{\label{2} (a)The potential energy as a function of $L$ at $T =0.1\mathrm{GeV}$ in the confined phase for the transverse case. The solid lines are the energies of $E_{Q\bar{Q}}$ and dashed lines are energies of $E_{Q\bar{q}}+E_{\bar{Q}q}$. (b)The dependence of running coupling constant $\alpha_{\mathrm{Q} \bar{Q}}$ on $L$ at $T =0.1\mathrm{GeV}$ in the confined phase for the transverse case. }
\end{figure*}

With these in hand, we turn to solve the problems numerically. First, the running coupling and string breaking of the quark-antiquark pair for the transverse case are shown in Fig.~\ref{2}. Note that we only draw the Coulombic part of Cornell potential. We show the energy $E_{Q\bar{Q}}$ of heavy meson and the energy $E_{Q\bar{q}}+E_{\bar{Q}q}$ of heavy-light meson at different $\omega$. The intersection of $E_{Q\bar{q}}+E_{\bar{Q}q}$ ($E_{Q\bar{q}}=E_{\bar{Q}q}$) and $E_{Q\bar{Q}}$ is where the string breaks. As the angular velocity increases, one can find that the string breaking distance increases. The string breaking happens at $L= 1.23\rm fm$, $L=1.24 \rm fm$ and $L=1.25\rm fm$ for $\omega= 0$, $\omega= 0.05\rm GeV$ and $\omega= 0.06\rm GeV$. Besides, it clearly shows that the maximum of running coupling decreases with the increase of angular velocity. However, the effect of rotation on the potential energy is not very significant in the transverse case.

\begin{figure*}
	\centering
	\includegraphics[width=17cm]{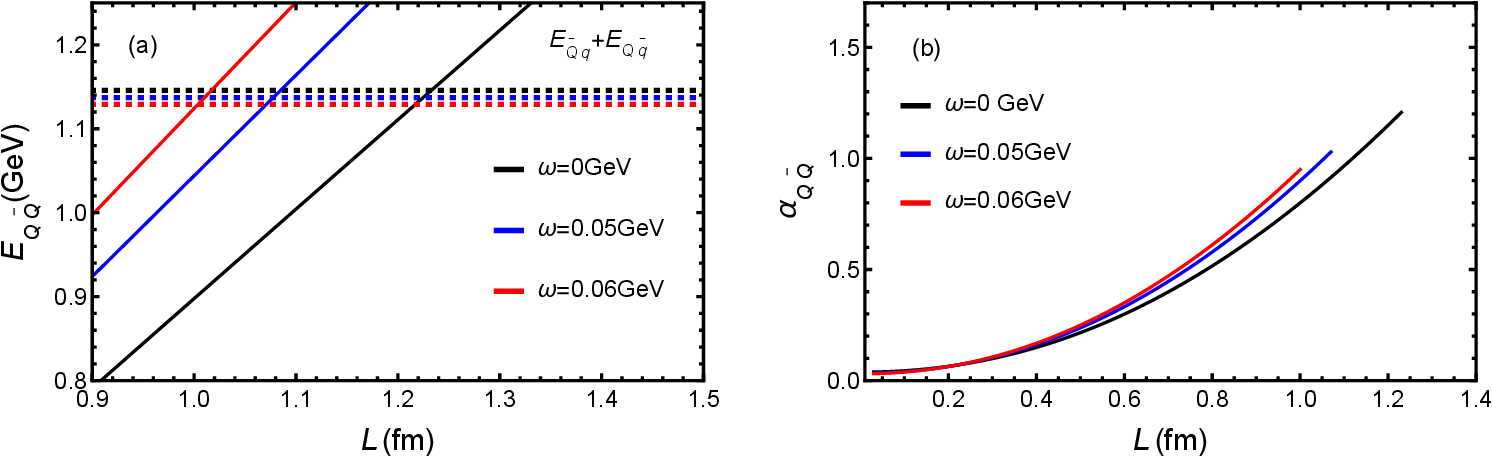}
	\caption{\label{3}(a)The potential energy as a function of $L$ at $T =0.1\mathrm{GeV}$ in the confined phase for the parallel case. The solid lines are the energies of $E_{Q\bar{Q}}$ and dashed lines are energies of $E_{Q\bar{q}}+E_{\bar{Q}q}$. (b)The dependence of running coupling constant $\alpha_{\mathrm{Q} \bar{Q}}$ on $L$ at $T =0.1\mathrm{GeV}$ in the confined phase for the parallel case. }
\end{figure*}

Then, we discuss the parallel case shown in Fig.~\ref{3}. The potential energy will increase with the increase of the $\omega$. Then, we found the string breaking happens at $L= 1.23\rm fm$, $L=1.07 \rm fm$ and $L=1\rm fm$ for $\omega= 0$, $\omega= 0.05\rm GeV$ and $\omega= 0.06\rm GeV$. The running coupling will reach the maximum at the string breaking distance which is a decreasing function of $\omega$. Therefore, we can conclude that the configuration of $\rm Q\bar{Q}$ becomes unstable in the parallel case. The effect of rotation in parallel case is more significant than that in the transverse case.

\section{Running coupling constant and free energy in deconfined phase}\label{sec:04}
\begin{figure*}
	\centering
	\includegraphics[width=17cm]{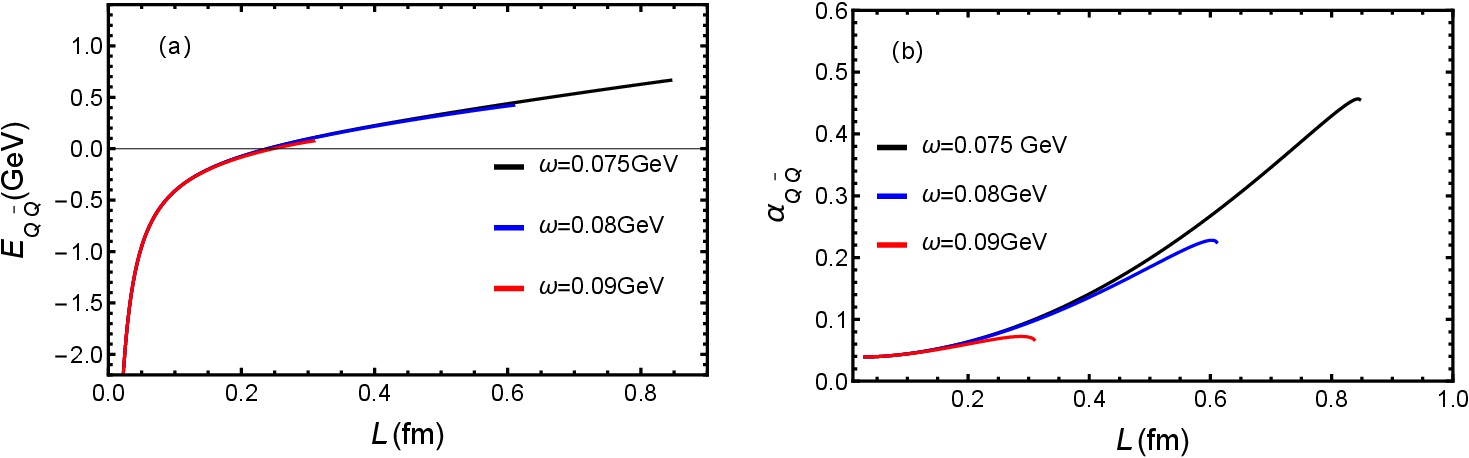}
	\caption{\label{5} (a)The potential energy as a function of $L$ at $T =0.1\mathrm{GeV}$ in the deconfined phase for the transverse case. (b)The dependence of running coupling constant $\alpha_{\mathrm{Q} \bar{Q}}$ on $L$ at $T =0.1\mathrm{GeV}$ in the deconfined phase for the transverse case.}
\end{figure*}

In our previous work \cite{Zhou:2021sdy}, one can find that the angular velocity can change the system from confinement phase to the deconfinement phase. In this section, we mainly concern the running coupling of $\rm Q\bar{Q}$ in the deconfined phase. We draw the running coupling in the deconfinement phase. Fig.\ref{5} shows that the screening distance is a decreasing function of $\omega$. The running coupling constant slowly increases to the maximum value and then decreases a little bit. The vanishing of running coupling constant means the dissolution of $\rm Q\bar{Q}$. With the increase of $\omega$, the maximum value of running coupling decreases.
\begin{figure*}
	\centering
	\includegraphics[width=17cm]{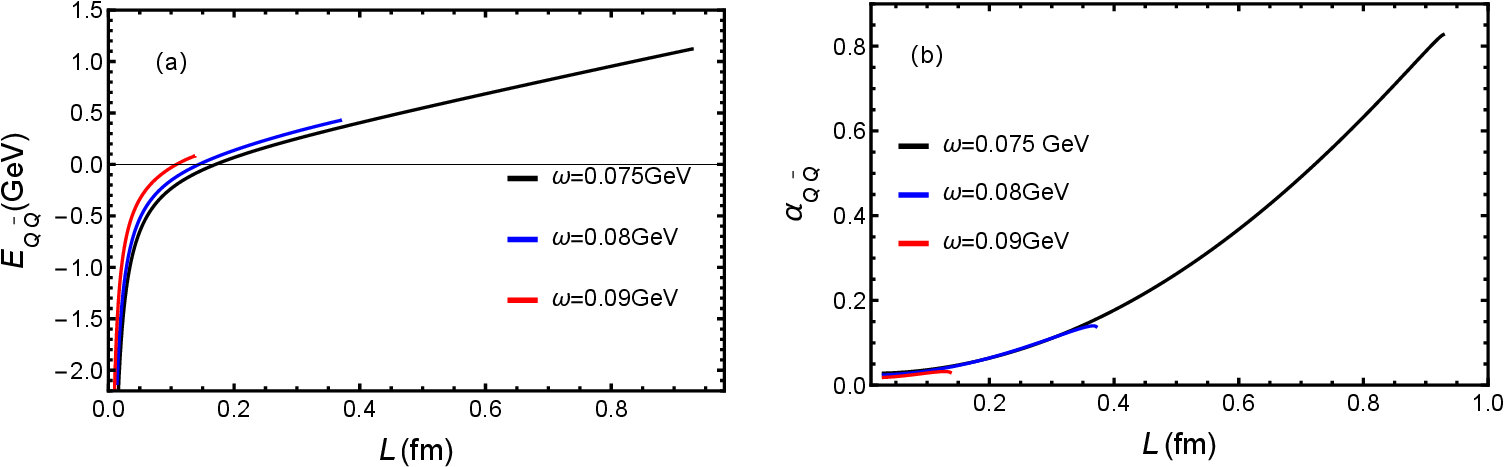}
	\caption{\label{6} (a)The potential energy as a function of $L$ at $T =0.1\mathrm{GeV}$ in the deconfined phase for the parallel case. (b)The dependence of running coupling constant $\alpha_{\mathrm{Q} \bar{Q}}$ on $L$ at $T =0.1\mathrm{GeV}$ in the deconfined phase for the parallel case. }
\end{figure*}
We also show the parallel case in Fig.\ref{6}. It is shown that the screening distance is smaller than that in the transverse case for a fixed $\omega$. Meanwhile, the maximum of the running coupling is smaller than that in the transverse case for a fixed $\omega$. We can conclude that $\rm Q\bar{Q}$ becomes easier to dissolve for the same angular velocity in the parallel case.

\section{Summary and conclusion}\label{sec:05}
In this work, we mainly study how the angular velocity affects the string breaking and running coupling constant of the quark-antiquark pair. First, we introduce a rotating black hole through Lorentz boost. Then, we calculate formulas of the running coupling and discuss the string breaking of the quark-antiquark pair. It is found that finite angular velocity can reduce the maximum value of running coupling constant. The transverse case has little influence on the string-breaking distance. The parallel case has a great influence on $\rm Q\bar{Q}$. The string-breaking distance will decrease with the increase of angular velocity in the parallel case. Also, the the maximum value of running coupling constant is decreasing at finite angular velocity. In the deconfined phase, running coupling constant has a maximum value beyond which the $\rm Q\bar{Q}$ will dissolve soon. The maximum value of the running coupling constant will decrease more quickly in the parallel case than that in the transverse case. Besides, due to the sea quark effect, many other decay modes like $\rm Q\bar{Q} \rightarrow Qqq + \bar{Q} q$ is deserved to be discussed in the future work.

\section*{Acknowledgments}\label{sec:06}
This work is partly supported by the National Natural Science Foundation of China under Grant Nos. 12005056 and 11947088, the Natural Science Foundation of Hunan Province, China(Grant No.2021JJ40020) and the Research Foundation of Education Bureau of Hunan Province, China under Grant Nos. 22B0788, 21B0402, and 19B100.

\vspace{10mm}

\vspace{10mm}
\bibliographystyle{elsarticle-num}
\biboptions{sort&compress}
\bibliography{reference}

\end{document}